\documentclass{aastex62}
\usepackage{natbib}
\usepackage{hyperref}
\usepackage{epstopdf}
\DeclareGraphicsExtensions{.ps}
\graphicspath{{./}{figures/}}
 
\begin{document}

\title{WISPR Imaging of a Pristine CME}

\correspondingauthor{Phillip Hess}
\email{phillip.hess@nrl.navy.mil}
\author{Phillip Hess}
\affiliation{U.S. Naval Research Laboratory, Washington, D.C.}

\author{Alexis P. Rouillard}
\affiliation{IRAP, Universit\'e Toulouse III - Paul Sabatier, CNRS, CNES, Toulouse, France}

\author{Athanasios Kouloumvakos}
\affiliation{IRAP, Universit\'e Toulouse III - Paul Sabatier, CNRS, CNES, Toulouse, France}

\author{Paulett C. Liewer}
\affiliation{Jet Propulsion Laboratory, California Institute of Technology, Pasadena, CA}

\author{Jie Zhang}
\affiliation{George Mason University, Fairfax, VA}

\author{Suman Dhakal}
\affiliation{George Mason University, Fairfax, VA}

\author{Guillermo Stenborg}
\affiliation{U.S. Naval Research Laboratory, Washington, D.C.}

\author{Robin C. Colaninno}
\affiliation{U.S. Naval Research Laboratory, Washington, D.C.}

\author{Russell A. Howard}
\affiliation{U.S. Naval Research Laboratory, Washington, D.C.}

\begin{abstract}The Wide-field Imager for Solar Probe (WISPR) on board the Parker Solar Probe (PSP) observed a CME on 2018 November 01, the first day of the initial PSP encounter. The speed of the CME, approximately 200-300 km s$^{-1}$ in the WISPR field of view, is typical of slow, streamer blowout CMEs. This event was also observed by the LASCO coronagraphs. WISPR and LASCO view remarkably similar structures that enable useful cross-comparison between the two data sets as well as stereoscopic imaging of the CME. Analysis is extended to lower heights by linking the white-light observations to EUV data from AIA, which reveal a structure that erupts more than a full day earlier before the CME finally gathers enough velocity to propagate outward. This EUV feature appears as a brightness enhancement in cooler temperatures such as 171 \AA, but as a cavity in nominal coronal temperatures such as 193 \AA. By comparing this circular, dark feature in 193 \AA \ to the dark, white-light cavity at the center of the eruption in WISPR and LASCO, it can be seen that this is one coherent structure that exists prior to the eruption in the low corona before entering the heliosphere and likely corresponds to the core of the magnetic flux rope. It is also believed that the relative weakness of the event contributed to the clarity of the flux rope in WISPR, as the CME did not experience impulsive forces or strong interaction with external structures that can lead to more complex structural evolution. 
\end{abstract}

\section{Introduction}
In 2018, the Wide-Field Imager for Solar Probe (WISPR; \citealt{wispr_2016}) on-board the Parker Solar Probe (PSP; \citealt{psp_2016}) became the most recent spacecraft to image the solar corona in white-light, and the first to do so within 0.3 AU. White-light observations of the corona date back to the 1970's, with missions such as OSO-7 \citep{tousey_1973}, Skylab \citep{macqueen_1974}, the Solar Maximum Mission (SMM; \citealt{macqueen_1980}), and P78-1 (Solwind; \citealt{sheeley_1980}).

For the last two decades, consistent imaging of the corona has been done by the Large Angle Spectrometric Coronagraph (LASCO; \citealt{lasco_1995}) from near the Sun-Earth line on-board the Solar and Heliospheric Observatory (SOHO). The Sun Earth Coronal Connection and Heliospheric Investigation (SECCHI; \citealt{secchi_2008}) on board the two STEREO spacecraft provided the first regular observations of the heliosphere from large angles away from the Earth, though still near 1 AU.

PSP is a unique and unprecedented mission for the heliophysics community, and is  structured differently from past missions in that it is not steadily taking data from a consistent distance. Instead the spacecraft is getting progressively closer to the Sun, with an ultimate minimum perihelion of 9.86 $R_{\Sun}$.  WISPR observes a fixed angular field of view of $13.5^{\circ}$ to $108.5^{\circ}$ with two detectors, WISPR-I (inner; $13.5^{\circ}$-$53^{\circ}$) and WISPR-O (outer; $50.5^{\circ}$-$108.5^{\circ}$) and observes the solar corona at lower heights as the mission progresses and the perihelion distance is reduced, allowing for the observation of coronal structures from closer distances than have ever been possible.

Due to the highly elliptical orbit of the spacecraft that is required to reach these low heights, the radial distance of the satellite to the Sun will also change significantly during each individual orbit. Unlike most previous heliospheric observing missions, PSP does not provide continuous observations. Instead, the mission is designed on an encounter basis, taking observations for about 10 days at a time, increasing in cadence as the spacecraft gets closer to perihelion. The first encounter began on 2018 November 01 at 53.8 $R_{\Sun}$, with a  perihelion on November 05 at 35.6 $R_{\Sun}$.

As a white-light instrument, WISPR is observing a combination of the stable, dust-based F-corona \citep{stenborg_2018} and the dynamic K-corona, consisting of photons scattered by electrons and comprised largely of solar outflows and transients \citep{howard_2009}. 

Coronal mass ejections (CMEs) are eruptions of magnetized plasma from the low corona, and are among the most notable and well studied solar transients because of the geomagnetic impacts they can cause at the Earth (\citealt{webb_2012} and references therein).

The frequency of CMEs is strongly correlated to the solar cycle (\citealt{hess_2017} and references therein). At solar maximum, it is common to see multiple CMEs a day erupting from all over the Sun, but at solar minimum there are far fewer eruptions and the CMEs tend to be weaker \citep{webb_1994}. Because PSP launched near solar minimum, it was unknown if WISPR would see any CMEs during its initial encounter. Fortunately, a CME entering the inner telescope on the first day of the encounter was obvious in the images.

The CME persisted long enough in the images that, before performing any detailed analysis on the event, it was obvious that this was a slower, less energetic eruption. Given the low speed and the proximity of the event to the  streamer structure that persists in the images throughout the first encounter, this event most likely belongs to the so-called 'streamer blowout' subset of CMEs \citep{Vourlidas_2018}, not to be confused with disconnection or 'pinch-off' eruptions formed at the top of the streamer cusp most commonly called streamer blobs \citep{Wang_1998}.

Streamer blowouts refer to a specific classification of events with a unique kinematic profile, featuring a gradual rise phase before the eruption \citep{srivastava_1999}, which is an obvious contrast with the faster, more impulsive CMEs that are often associated with flares \citep{zhang_2004}. These events originate from the streamer belt, but often feature cavities that can be tracked from underneath the streamer until exiting through the streamer cusp \citep{sheeley_1997}. The final speeds of these events tend to reach and stay at the ambient solar wind velocity, so these events are often considered to be tracers of the acceleration and speed of the solar wind \citep{srivastava_2000,sheeley_1999}.

\begin{figure}
    \centering
    \includegraphics[scale=0.5, trim=50 100 50 50, clip]{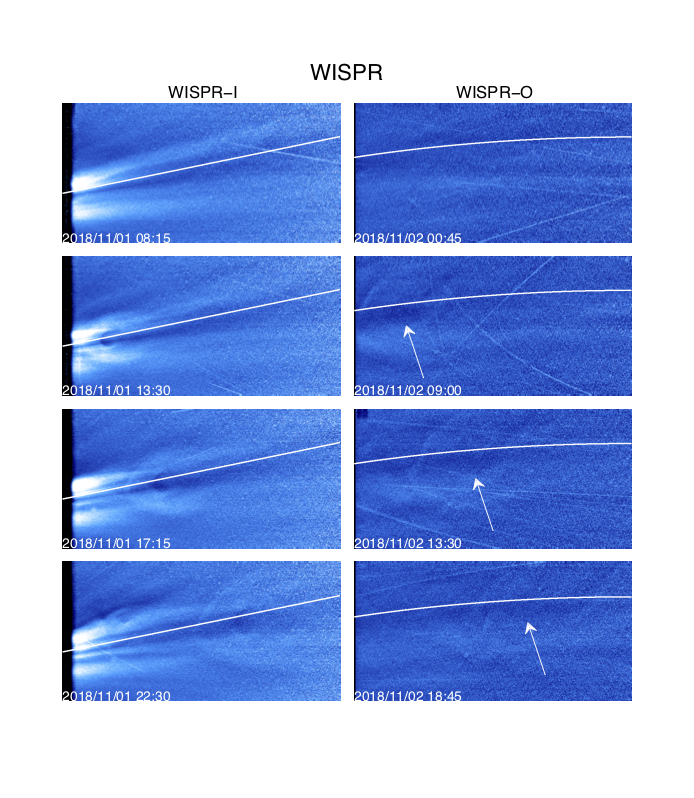}
    \caption{The CME in both WISPR-I (Left) and WISPR-O (Right). The images from both detectors are cropped to 800x400 pixels. In Detector 2, the white arrows were added to point out where the core of the CME is located. The first image in a each sequence is a pre-event image. The solar equatorial plane has been added as a white line in each image for reference. The bend in the equatorial plane in WISPR-O is a result of the distortion of the detector.}
    \label{fig_wispr}
\end{figure}

In this paper, we will present the remote sensing observations of this CME in WISPR, as well as other instruments that provide context for the event. This includes the two LASCO coronagraphs as well as extreme ultraviolet (EUV) data from the Atmospheric Imaging Assembly (AIA; \citealt{AIA_2012}) on-board the Solar Dynamics Observatory (SDO). Using all of these data, we will present the complete evolution of the eruption, from the initial rising of the ejecta in the low corona until it exits the outer WISPR telescope. A second paper will focus on the modeling and a theoretical analysis of this event \citep{Rouillard_2019}.

\begin{figure}
    \centering
    \includegraphics[scale=2]{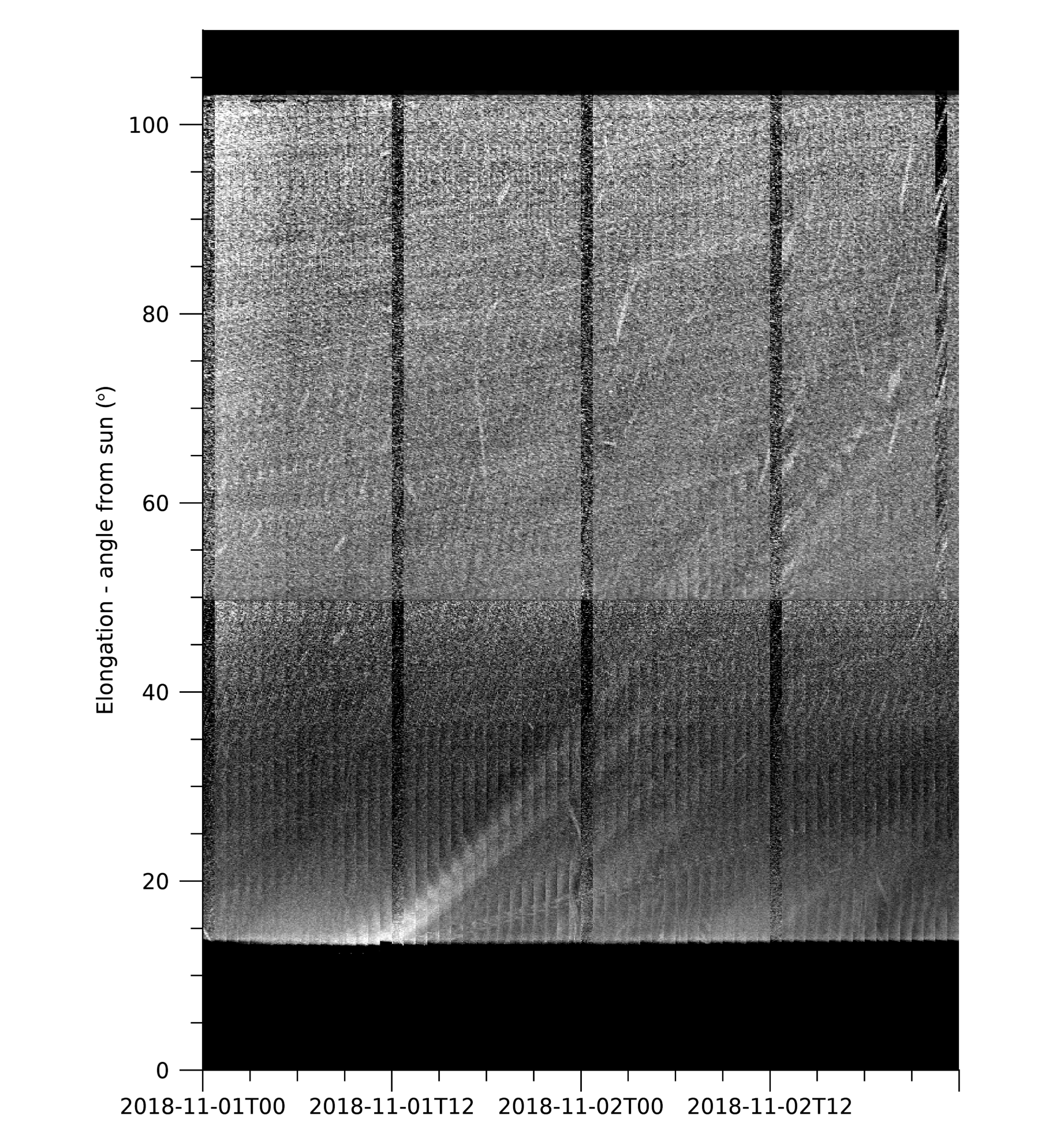}
    \caption{Height-time plot or J-map of the CME in both WISPR Detectors taken from the spacecraft orbit plane.}
    \label{fig_wisprjmap}
\end{figure}

\section{White-Light Observations}
Obtaining a useful signal in physically meaningful units from WISPR data is not a trivial task. To process the raw data, the Level-1 FITS files containing brightness measurements in DN must be normalized for exposure time, and then corrected for the vignetting effects of the detector. To then create Level-2 data in physical units, a calibration constant is then used to convert DN s$^{-1}$ into mean solar brightness (MSB) units. The vignetting function and calibration constant were both initially determined during pre-flight calibration of the instrument but have since been modified based on stellar photometry, similar to methods employed on past coronagraphs \citep{morrill_2006, thernisien_2006} and heliospheric imagers \citep{bewsher_2012, tappin_2015}. However this work is still preliminary, and after more orbits the calibrations will continue to be modified.

With the data in MSB units and calibration effects corrected, the signal in the image is dominated by the F-corona. Given the rapid movement of the spacecraft, combining many images to determine a stable background will not effectively remove the F-corona in WISPR. Adapting a technique developed by \citep{stenborg_2018} on SECCHI HI-1 data, a background removal technique has been applied to WISPR images allowing for transient K-corona features to be highlighted. Like the calibrations, this technique is preliminary and will continue to be improved as more data are collected. This technique was applied to all the WISPR images presented in the paper.

On 2018 November 01, PSP entered its first encounter with the solar corona. In the first images taken by WISPR during the encounter a streamer was plainly visible in the inner detector close to the solar equatorial plane. This streamer can be seen in each detector in Figure \ref{fig_wispr}. The streamer appears tilted above the mid-plane of the image due only to the pointing and roll of the instrument.

The CME came from the same position angle as the steamer and  entered the FOV by 11:15 UT. However, it is not until 12:45 UT that the CME can be clearly distinguished from the streamer cusp. The CME then enters the WISPR-O FOV around 03:45 UT on November 02, growing increasingly faint as it propagates before likely leaving the FOV around 11:15 UT on November 03. 

The CME in both detectors is presented in Figure \ref{fig_wispr}. While the CME is difficult to isolate clearly in the outer detector, there are some common characteristics of the event present in each detector. The CME has a strong 'V' shape at the trailing edge. In front of this feature, which is present as a brightness enhancement, there is a coherent circular dark feature. This dark cavity-like structure persists until the CME is no longer visible in WISPR-O.

The CME was faint and relatively slow. Given that the total FOV of WISPR covers about 80 $R_{\Sun}$ (an approximate value, given the changing radial coverage of the instruments as the spacecraft gets closer to the Sun), the CME taking two full days to traverse through the full FOV is indicative of a speed of $\sim$300 km s$^{-1}$. This low speed, combined with the co-spatial streamer visible in WISPR-I, indicated that the event was most likely a steamer blowout.

\begin{figure}
    \centering
    \includegraphics[scale=0.5, trim=0 0 0 37, clip]{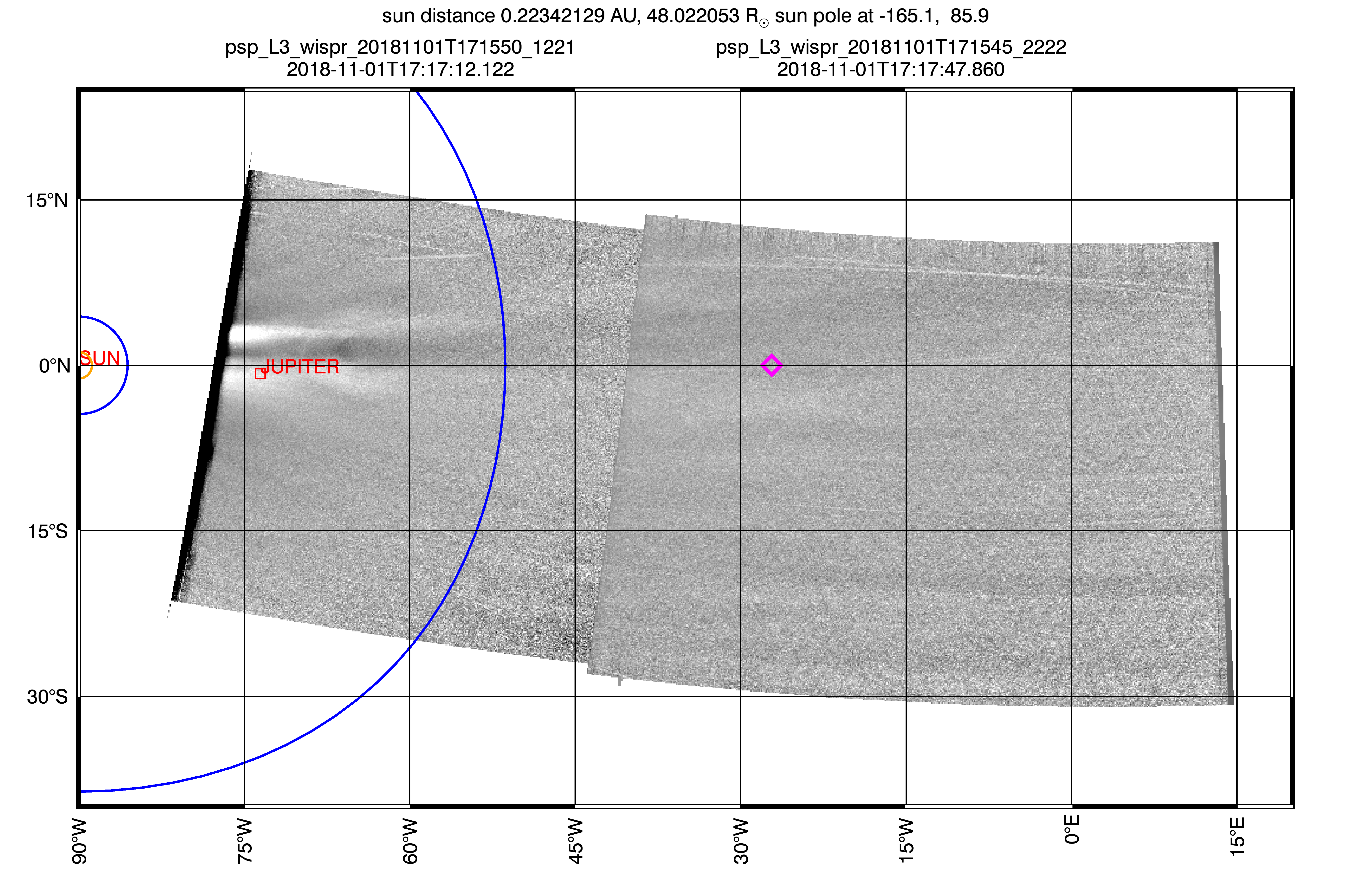}
    \caption{Observation of the CME in WISPR, in a heliocentric coordinate system with 0$^{\circ}$ latitude at the PSP orbital plane (see text). The blue circles indicate the approximate position of the C3 field of view based on a projection of the Thomson sphere onto the image plane and ignoring the longitudinal separation of the spacecraft. The pink diamond represents position of the velocity vector of the spacecraft on the image plane. The position of Jupiter is marked as well.}
    \label{fig_fov}
\end{figure}

Figure \ref{fig_wisprjmap} shows a plot of elongation (angle from the Sun) vs. time (referred to as a J-map \citep{sheeley_1999}) from WISPR observations on 2018 November 01-02, where the elongation is measured in the plane of PSP's orbit. The J-map was made by taking a strip of pixels several degrees wide at the PSP orbit plane across the full elongation range of each projected image in the sequence and stacking them vertically to form Figure \ref{fig_wisprjmap}. To make the J-map, the sequence of WISPR images for both telescopes were first re-projected into the PSP orbit plane frame of reference. This frame is defined by two vectors that define the orbit plane: the Sun-PSP vector and PSP's velocity vector. The origin of this coordinate system was chosen to be in the orbit plane at the direction perpendicular to the vector from Sun to PSP, putting the Sun at a constant 90$^{\circ}$ West.  A sample projected image from November 01 at 17:17 UT in this frame of reference is shown in Figure \ref{fig_fov}. The track of the flux rope's dark cavity can also be seen in Figure \ref{fig_wisprjmap}, crossing from WISPR-I to WISPR-O FOV at roughly 06:00 UT on November 02.

\begin{figure}
    \centering
    \includegraphics[scale=0.5, trim=50 100 50 50, clip]{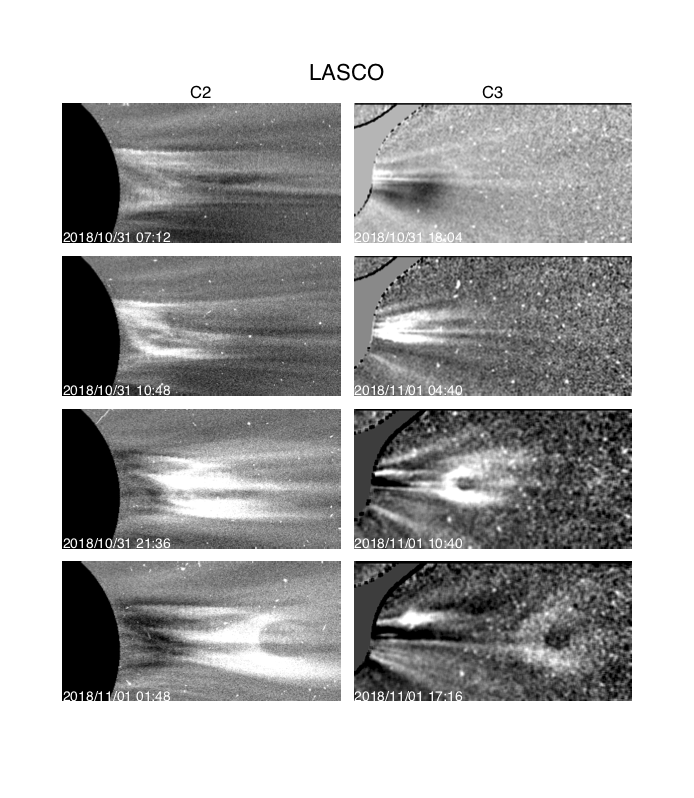}
    \caption{The CME as observed in LASCO C2 and C3 from the October 31 into November 01. The images from both detectors are cropped to 300x200 pixels. The first frame in each series shows a pre-event image.}
    \label{fig_lasco}
\end{figure}

It appears as though the CME is accelerating throughout the FOV of each detector. This is likely an effect of the FOV changing as the radial distance between the spacecraft and the Sun changes during the orbit. At the beginning of the J-map in Figure \ref{fig_wisprjmap} the spacecraft is 51.6 $R_{\Sun}$  from the Sun and by the end of the J-map, it is at 43.0 $R_{\Sun}$. While this change may seem relatively minor, it can have significant impacts on the conversion between elongation and radial distance. Converting 50$^{\circ}$ elongation on a plane 90$^{\circ}$ from WISPR into a radial distance, the calculated height reduces by $\sim$10 $R_{\Sun}$ from the beginning of Figure \ref{fig_wisprjmap} to the end. 

There is also the possibility that the spacecraft could be getting closer to the CME as it propagates, causing a projection effect that would lead to artificial enhancements to both the bulk and expansion velocities of the CME as the distance between the CME and the satellite decreases. Getting reliable height and velocity measurements from a more stationary instrument is possible and has been done throughout the STEREO mission with various different methods \citep{Barnes_2019, mishra_2014}. However, every method does require some simplifications and assumptions, and the extra uncertainties imposed by WISPR make estimating a velocity from J-map significantly more complex \citep{Liewer_2019}.

\begin{figure}
    \centering
    \includegraphics[scale=0.45, trim=80 0 0 0, clip]{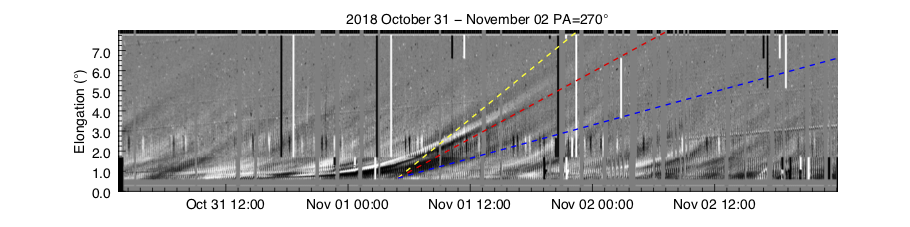}
    \caption{J-map from LASCO beginning on October 31, combining the C2 and C3 data at position angle 270$^{\circ}$ (roughly 0$^{\circ}$ heliographic latitude). The CME is both the bright feature as well as the dark void in front, which correspond to the trailing edge and cavity, respectively. For reference, lines with slopes representing 100 km s$^{-1}$ (blue), 200 km s$^{-1}$ (red) and 300 km s$^{-1}$ (yellow) velocities have been included.}
    \label{fig_jmap}
\end{figure}

To confirm the information obtained from the WISPR images the data was compared to that of other instruments to observe the initial eruption and the evolution of the CME. At the time of this eruption, WISPR was about 40$^{\circ}$ West of the Sun-Earth line and looking back towards the Sun, relative to the Earth. Because of location of the satellite, anything imaged in WISPR was likely to be on the back side of the Sun relative to \textit{STEREO}-A, as the CME was a slow and dim event propagating right behind the occulter, this meant that SECCHI data would not be particularly useful.

The LASCO coronagraphs, however, are always positioned to observe activity on the western limb of the Sun. Figure \ref{fig_fov} compares the WISPR FOV to that of SOHO/LASCO/C3, which is indicated by the blue circle. The C3 field-of-view extends to about 30 $R_{\Sun}$. The pink diamond in Figure \ref{fig_fov} shows the direction of PSP's velocity vector at this time. The angle between the directions of the velocity vector and the origin reflects the ellipticity of the PSP orbit; a circular orbit would have the velocity vector perpendicular to the Sun-SC line, e.g., the direction of the line-of-sight at the origin of the PSP orbit plane frame.  PSP's orbit plane is Venus' orbital plane (because PSP uses Venus flybys to shrink its orbit), which is inclined by about 3.4$^{\circ}$ from the solar equatorial plane.

Because the spacecraft are observing from different positions and the range of the WISPR detectors will change as the CME gets closer to the Sun, any attempt to compare the fields of view is just an approximation. However, based on Figure \ref{fig_fov}, it can be expected that anything that is seen by the inner half of WISPR-I detector while observing close to the western limb should be seen co-temporally in C3.

A few events around the CME were visible in LASCO during the WISPR observations, with one clear choice for a counterpart to the WISPR CME. On November 01, there was a CME observed by C3 that was from a streamer in the solar equatorial plane. This CME was slow, and had a dark, circular cavity in front of a bright trailing edge that came to a cusp at the back end of the CME. A series of snapshots from both LASCO coronagraphs from October 31 through November 02 is shown in Figure \ref{fig_lasco}.

The C3 image taken at 17:16 UT overlaps with the WISPR-I image from Figure \ref{fig_wispr} taken at 17:15 UT, and the structure looks similar in each detector, though obviously the resolution is better in WISPR. In LASCO, the CME appears much more separated from the streamer than at the same time in WISPR, indicating that the part of the streamer that is well observed in WISPR is likely behind the CME and appears close only because of projection effects. 

This same CME was also visible in C2.  The claw-like structure seen in both WISPR and C3 first becomes visible about halfway through the C2 field of view. Before that, it is difficult to separate the CME from the streamer that is present. As seen in Figure \ref{fig_lasco} at 10:48 UT on October 31, there does appear to be a a circular, dark feature that could correspond to the cavity observed later. This feature is not radially oriented, and its motion is highly non-radial. This indicates an eruption from a higher latitude being pushed by the overlying magnetic field to the equator until the flux rope reaches the open field lines of the streamer and can then lift off, behavior that was also observed in STEREO \citep{Liewer_2015}.

To demonstrate the kinematics of this event, a LASCO J-map is shown in Figure \ref{fig_jmap}. The brightest feature in the J-map corresponds to the trailing edge of the CME, and there is a noticeable expanding, dark feature in front of the trailing edge in C3. The CME rises very gradually between October 31 and November 01, staying within 1$^{\circ}$ in elongation of the solar surface for nearly 8 hours after the CME first becomes distinguishable in C2. Eventually, the CME picks up speed after it leaves the C2 field of view and propagates approximately linearly through the C3 field of view. The speed determined by a simple estimation of the slope of the density enhancement supports a roughly 300 km s$^{-1}$ velocity for the event. 

\begin{figure}
    \centering
    \includegraphics[scale=0.55]{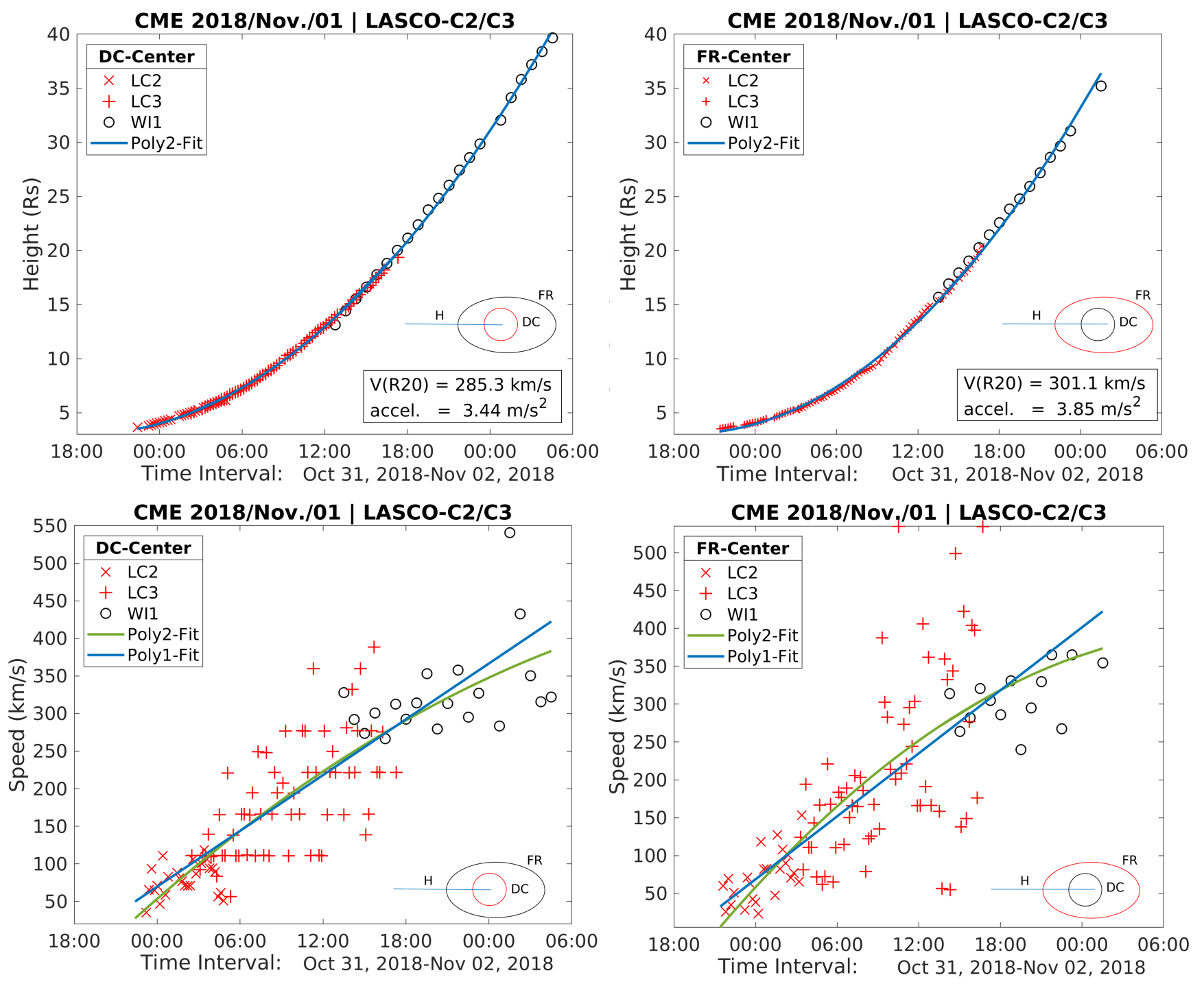}
    \caption{Top: Kinematics of the cavity (left panel) and flux rope (right panel) from combined observations of WISPR and SOHO/LASCO. In both panels, red and blue markers are used when the features height is determined from LASCO or WISPR images, respectively. A second order polynomial function was used to determine the speed and the acceleration of the cavity and flux rope at 20 $R_\Sun$. Each feature is measured from its geometric center, based on the circular or elliptical feature as depicted in the schematic at the bottom right corner of the plot. Bottom: Velocity profiles of each feature. The points in these plots are the difference of the height measurements in the top row. The blue line is the analytical derivative of the polynomial fit to the height measurements. The green line is a second order polynomial fit to the velocity points.}
    \label{fig_evo}
\end{figure}

This approximate result can be confirmed with a more complex kinematic analysis of the bright envelope or claw-like feature, hereafter referred to as the flux rope, and the dark cavity using both WISPR and LASCO. Figure \ref{fig_evo} shows the resulting height versus time plots for the flux rope and the dark cavity (right and left panel, respectively). To perform the measurements an ellipse was fit to the outermost edges of the two features that appear in the images. The kinematics are then determined by the temporal evolution of the centroid of the fitted ellipses. The height-time measurements using LASCO images and those made by WISPR observations seem to agree well. The similarity of the images indicate that, despite an angular separation of approximately 40$^{\circ}$, the FOVs for WISPR and LASCO overlap and allow for a consistent comparison for this CME.

These plots again show a very slow rise of the CME from the low to the upper corona. The speed of dark cavity is roughly similar with the estimated speed of the flux rope. At 20 $R_\sun$ the speed of the dark cavity is estimated to be 285 km s$^{-1}$ and at the same height the flux rope speed of 300 km s$^{-1}$, based on second order polynomial fits to the height measurements. The acceleration happens in two stages, an initial, low acceleration in the corona before a higher acceleration to a potential slow solar wind speed. Both the cavity and flux rope features undergo the same gradual acceleration profiles around 3.5 km s$^{-1}$${^2}$.

\section{EUV Observations} \label{sec_AIA}

\begin{figure}
    \centering
    \includegraphics[scale=0.7, trim=0 0 0 0, clip]{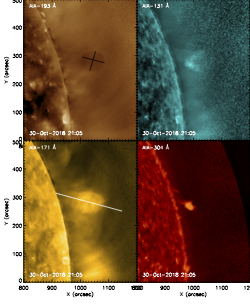}
    \caption{AIA images during the eruption on October 30. Different wavelengths enhance different elements of the CME including the dark, circular cavity (193 \AA; 1.6 MK and 20 MK), the bright trailing edge (131 \AA; 0.4 MK and 10 MK), a bright blob that is co-spatial with the cavity (171 \AA; 0.6 MK) and the prominence at the base of the eruption (304 \AA; 0.05 MK). The black line in the 193 \AA \ frame was used to calculate the size of the cavity. The white line in the 171 \AA \ frame is the approximate direction of motion and was used to measure the height and calculate the velocity of the cavity in AIA.}
    \label{fig_aia}
\end{figure}

On 2018 October 30 at 16:00 UT, a circular cavity was observed at a heliographic latitude of 30$^{\circ}$ in SDO/AIA. An associated prominence was observed $\sim$37.5 Mm above the solar limb. Around 19:00 UT the cavity started to rise slowly and non-radially, towards the solar equator. Figure \ref{fig_aia} shows the eruption of the cavity in different AIA passbands at 21:05 UT. The erupting cavity is observed as a bright blob in AIA 171 \AA$\sim$(0.6 MK) and the corresponding feature was observed as dark cavity in the AIA 193 \AA \ (combination of 1.6 MK and 20 MK) passband. The 131 \AA \ (combination of 0.4 MK and 10 MK) passband shows a bright trailing edge with a remarkably similar shape to what is observed in both WISPR and C3. Considering the lack of hot plasma in other wavelengths, it is likely this comes from the cooler ions observed in 131 \AA. Also, the corresponding prominence was observed to rise in the cooler AIA passband 304 \AA$\sim$(0.05 MK). 

\begin{figure}
    \centering
    \includegraphics[scale=0.5, trim=0 00 0 0, clip]{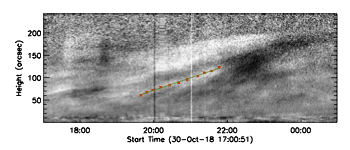}
    \caption{The height-time plot showing the eruption of the cavity in AIA 171 \AA. The position of the slice is shown in the lower-left panel of the Figure \ref{fig_aia}. The red asterisks mark the positions of the lower edge of the erupting cavity and the linearly fit to the points is in green.}
    \label{fig_slice}
\end{figure}

The size of the cavity in AIA 193 \AA, calculated along the major axis of the feature (see the black line in the top left panel Figure \ref{fig_aia}), was 59.2 Mm at 21:05 UT. The rising motion of the cavity was slow. To study the rising motion of the erupting cavity quantitatively a slit was made in running difference images from AIA 171 \AA \ to create a height-time plot similar to the white light J-maps (see the Figure \ref{fig_slice}). The position of the slit is shown by a white line in the lower-left panel of the Figure \ref{fig_aia}. While the velocity of the leading edge is difficult to determine as it leaves the field of view a few hours after the CME is first visible, the trailing edge can be tracked until the CME completely exits AIA. The rise of the trailing edge of the cavity is shown as red asterisk in the Figure \ref{fig_slice}. Using a linear fit on the trailing edge, the velocity was determined to be 5.5 km s$^{-1}$ (shown with the green line in Figure \ref{fig_slice}).

\begin{figure}
    \centering
    \includegraphics[scale=0.5, trim=0 00 0 0, clip]{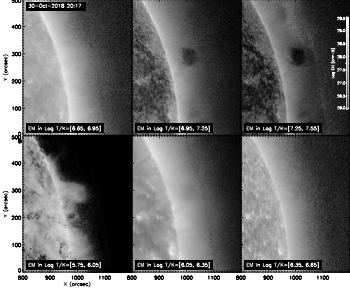}
    \caption{DEM maps during the eruption of the cavity on 2018 October, 30 at 20:17 UT by the sparse inversion method. Each panel shows the total EM at a different temperature range.}
    \label{fig_em}
\end{figure}

The temperature properties of the erupting cavity were investigated using the differential emission measure (DEM) method by employing the sparse inversion code \citep{Cheung_2015}. Figure \ref{fig_em} shows the emission measure (EM) map at different temperature ranges during the eruption of the cavity. Since the cavity was lifting off very slowly, six AIA passbands (94, 131, 171, 193, 211 and 335 \AA) were used in the time range of 20:12 UT to 20:17 UT (selecting the data with minimum noise) to calculate the EM. The EM maps suggest that the temperature of the erupting cavity was $\leq$ 1 MK.

There is a candidate filament at the proper longitude visible in AIA in the days leading up to the eruption, which is likely the same filament seen to erupt in the 304 \AA \ passband on October 30. It is not immediately apparent what caused the eruption. The filament is still relatively small, and there is nothing in the other wavelengths to indicate a strong eruption.

The lack of a striking low coronal signature is not surprising, considering the speed of the rising motion in AIA is just 5.5 km s$^{-1}$, there would not likely be a strong flare impulsively providing energy to the CME. Instead this is most likely an eruption of a pre-existing flux rope gradually losing equilibrium, possibly due to a small eruption that is visible just to the south of the cavity in 193 \AA \ in the hours before the eruption. This could have perturbed the overlying field just enough on the southern side to cause the flux rope to begin to rise. 

Because the field above the CME would have only been affected on one side, the still-strong magnetic field to the north of the CME deflects the flux rope to the south as it rises, causing the non-radial motion of the CME in AIA. Once the CME reaches the current sheet, it is again deflected, this time radially with the outflowing solar wind. The cavity, maintained by the strength of the magnetic field remains in tact while expanding as the ambient density around it drops.

The entire flux rope undergoes a similar expansion, but the brightness at the trailing end of the flux rope remains much much brighter than that of the leading edge throughout the propagation. It is likely that this CME is accelerated by the solar wind, and the compression of plasma on the back of the CME is likely why the CME is so much brighter on that side. 

\section{Summary}
On the first day of the first PSP encounter, WISPR imaged its first CME. By working backwards through multiple data sets in different observers, the entire evolution and morphology of this event can be pieced together. A structure existed in the low corona for an unknown length of time. In hotter wavelengths the structure appears as a cavity while appearing as a density enhancement in the cooler 171 \AA \ images. This indicates a cooler structure, confirmed by DEM analysis that indicates a temperature around 1 MK. At any temperature, the well defined circular structure suggests the structure is likely a flux rope.

Shortly after rotating onto the limb of AIA, something, most likely another nearby, faint eruption, weakens the overlying magnetic field and destabilizes the flux rope, causing it to rise gradually and non-radially toward the equator at a speed around 5.5 km s$^{-1}$. Over the course of nearly an entire day, the flux rope continues to slowly rise through the C2 field of view before it reaches the open field region and and begins to accelerate before crossing into C3. In the C3 FOV the acceleration declines until the CME is travelling at a fairly constant speed. The cavity at the center is also still visible, and can be seen to expand as the CME propagates.

Eventually, the CME, including the cavity at the center, enters the WISPR field of view. It propagates directly along the streamer belt lying over the equatorial plane. The J-map made from WISPR data shows signs of an apparent acceleration well into the WISPR-O field of view. It's unlikely that this CME would continue to accelerate at such a large distance, and instead this is most likely the result of the spacecraft getting closer to the CME as it propagates. 

A number of factors were aligned for this event to be well observed by multiple satellites. If the CME had gone off just a day earlier or a day later, the cavity would not have rotated onto the limb in AIA to be as clearly observed. LASCO would likewise have most likely missed the cavity and this event would have been just a typical slow streamer blowout. 

This is also indicates the insights that can be gained from WISPR data. Streamer blowouts are common events that are usually slow, poorly defined in 1 AU observations and overlooked in many studies. This event in particular was difficult to track through the C3 field of view, and so observations in the past for this event would have largely been restricted to about $20 R_{\Sun}$. However, with WISPR we were able to observe, track and measure a clear circular cavity indicative of a flux rope core from its initial eruption in the low corona out to nearly half an AU. 

The observation of this core is extremely valuable, as is the chance to study a flux rope with such a clear structure. Because this CME was not initiated by a large, impulsive energy release manifested as a flare and did not violently exit the corona, there were fewer strong forces contributing to a possible erosion or distortion of the flux rope. Also, because of the relatively weak expansion of the flux rope, the CME structure had a smaller spatial extent in the corona and heliosphere and was therefore less likely to interact with something on its flank (i.e. a streamer). By studying events such as this at a closer range with better resolution, it will be possible to see how a more idealized flux rope propagates through the heliosphere - an important test for theory and models, as is demonstrated in \citet{Rouillard_2019}.

\acknowledgments
The authors thank the referee for suggestions that improved the clarity of this manuscript. We acknowledge the work of the PSP operations team. The SOHO/LASCO data used here are produced by a consortium of the Naval Research Laboratory (USA), Max-Planck-Institut fuer Aeronomie (Germany), Laboratoire d'Astronomie (France), and the University of Birmingham (UK). SOHO is a project of international cooperation between ESA and NASA. The AIA data are courtesy of SDO (NASA) and the AIA consortium. This work was completed while PH was supported by the Jerome and Isabella Karle Distinguished Scholar Fellowship at NRL. PH, GS, RCC and RAH also acknowledge support from the NASA PSP program office.  AK acknowledges financial support from the ANR project SLOW{\_}\,SOURCE (ANR-18-ERC1-0006-01), COROSHOCK (ANR-17-CE31-0006-01), and FP7 HELCATS project \url{https://www.helcats-fp7.eu/} under the FP7 EU contract number 606692. The work of APR was funded by the ERC SLOW{\_}\,SOURCE project (SLOW{\_}\,SOURCE - DLV-819189). The work of PCL was conducted at the Jet Propulsion Laboratory, California Institute of Technology under a contract from NASA. JZ is supported by NASA grant NNH17ZDA001N-HSWO2R. SD is supported by the DKIST Ambassador program provided by the National Solar Observatory, a facility of the National Science Foundation, operated under Cooperative Support Agreement number AST-1400405. The authors thank Jeffrey R. Hall and Paulo Penteado for assistance in processing the data.

\bibliographystyle{apj}
\bibliography{wisprref}

\end{document}